\documentclass[prl,twocolumn]{revtex4}  

\usepackage{graphicx}
\usepackage{mathptmx}  

\begin{document}

\newcommand{\isotope}[2]{\mbox{$^{#2}\mbox{#1}$}}
\newcommand{\chilambda}{\ensuremath{\chi^2_{\lambda,p}}}

\noindent \textbf{Comment on ``Nuclear Emissions During Self-Nucleated
   Acoustic Cavitation''}

\vspace{0.3cm}

In a recent Letter \cite{taleyarkhan2006}, Taleyarkhan and coauthors
claim to observe DD fusion produced by acoustic cavitation.  Among
other evidence, they provide a proton recoil spectrum that they
interpret as arising from 2.45~MeV DD fusion neutrons.  My analysis
concludes the spectrum is inconsistent with 2.45~MeV neutrons, cosmic
background, or a \isotope{Pu}{239}Be~source, but it is consistent with a
\isotope{Cf}{252}~source.

\begin{figure}[h]
\centering
\includegraphics[width=3.375in]{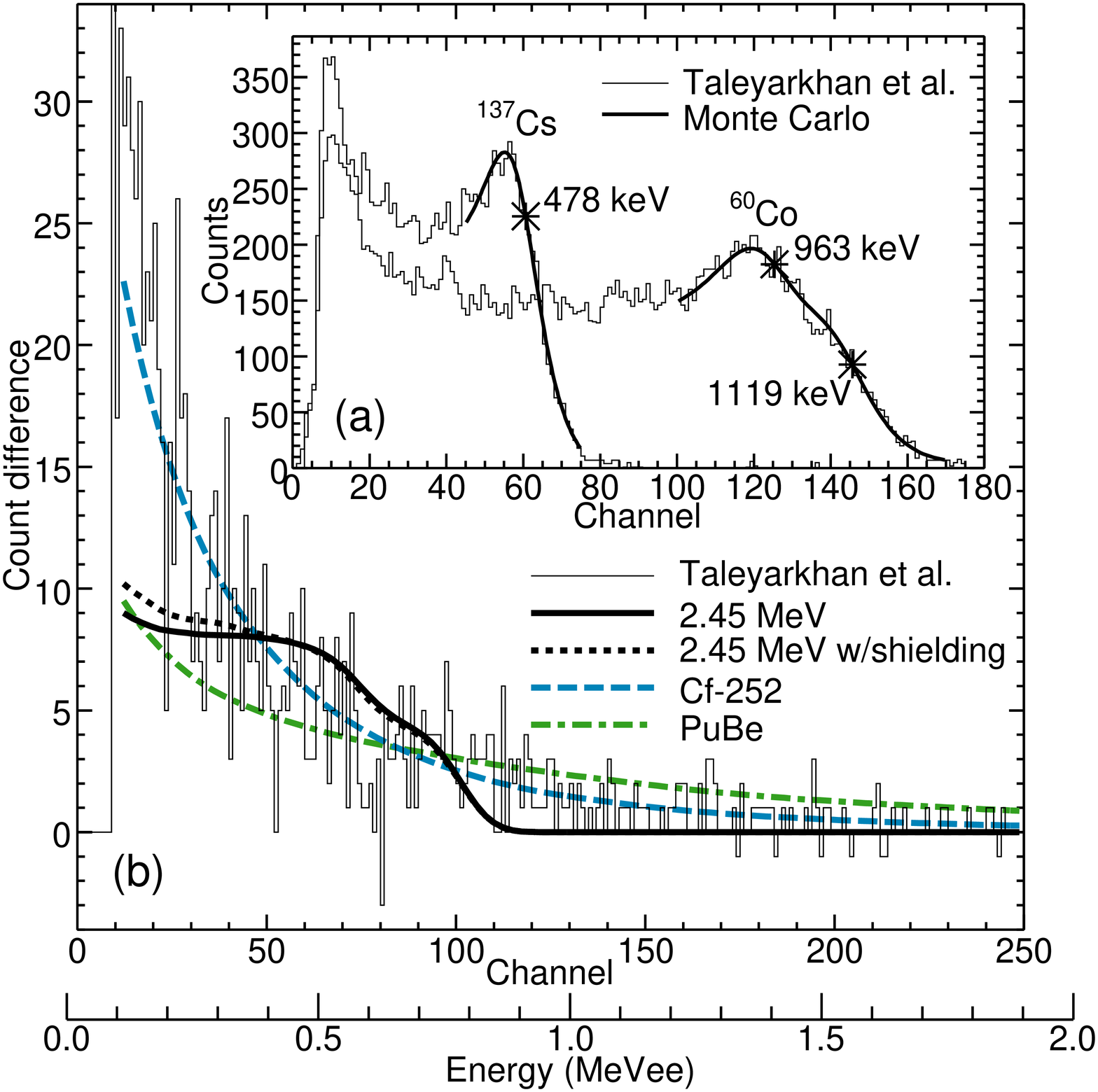}
\caption{
(color online) Analysis of Taleyarkhan and coauthors' liquid
scintillator data.
(a) Fitting the measured Compton edges of calibration $\gamma$ sources
    to simulated electron recoil spectra determines the detector's energy
    scale and resolution.
(b) Simulated proton recoil spectra of various candidate neutron sources
    shown fit to data.}
\label{fig:analysis}
\end{figure}

Figure~\ref{fig:analysis}(a) shows the detector's pulse height spectra
of two calibration $\gamma$ sources, as extracted from Fig.~8 of the
Letter's supplement \cite{taleyarkhan2006supp}.  I use \textsc{Geant4}
\cite{agostinelli2003} to simulate the detector's electron recoil
spectra, which are then convolved with a gaussian resolution function
and scaled to fit the measured spectra \cite{dietze1982}.  The two
fits, showing excellent agreement with the data, validate the method
and provide parameters for the detector's light output function $L = c
\; (E - E_0)$ and resolution \cite{birks1964} $\eta^2 = \alpha +
\beta/E$.

As described in the supplementary methods \cite{suppMethods}, I
simulate proton recoil spectra for the four separate cases.  In the
two limiting cases of 2.45~MeV neutron emission---no shielding and
heavy shielding---the detector is placed 30~cm from the quartz flask
containing the cavitation fluid.  The two radioisotope simulations
assume there are no intervening scattering materials.  These
techniques were used to accurately model a DD fusion proton recoil
spectrum in Ref.~\cite{naranjo2005}.

Figure~\ref{fig:analysis}(b) shows the simulated spectra, fit to data
extracted from Fig.~4 of the Letter.  As described in the
supplementary methods, the fit is performed simultaneously over the
raw cavitation `on' and cavitation `off' data, extracted from
Fig.~9(b) of the supplement.  The \chilambda~variable of
Ref.~\cite{baker1984} determines both the best fit parameters and the
goodness-of-fit.

The fit results, summarized in Table~\ref{table:analysis}, show the
data are statistically consistent with \isotope{Cf}{252}, since the
observed value of \chilambda~is within one standard deviation of the
mean.  In contrast, the observed values of
\chilambda~for the remaining cases are more than five
standard deviations beyond the mean, and, consequently, the data are
statistically inconsistent with DD fusion or a PuBe source.

\begin{table}[h]
\caption{Results of fit to simulation.  For each fit, numerical
sampling determines the distribution of goodness-of-fit variable
\chilambda.  Then, the number of standard deviations from the mean for
the observed value of \chilambda is reported as a Z-value.  See Ref.
\cite{suppMethods} for details.}
\begin{ruledtabular}
\begin{tabular}{lccc}
                        & \chilambda & Z-value \\ \hline
  2.45 MeV              &  653       & 5.9     \\
  2.45 MeV w/ shielding &  637       & 5.5     \\
  Cf-252                &  432       & -0.45   \\
  PuBe                  &  621       & 5.9     \\
\end{tabular}
\end{ruledtabular}
\label{table:analysis}
\end{table}

Comparing the shapes of the spectra in Fig.~9(b) of
Ref.~\cite{taleyarkhan2006supp} rules out the possibility of
cavitation `on' runs being longer than cavitation `off' runs.  Calling
channels ten and below the `peak' and channels eleven and above the
`tail', the ratio of tail to peak counts with cavitation off is
$291/764=0.38$.  When cavitation is on, the tail becomes more
pronounced so that the ratio is $1216/835=1.5$.

\acknowledgments{I thank S.~Putterman for valuable discussions.
This work is supported by DARPA.}

\vspace{0.3cm}

\small
\noindent B. Naranjo \\
\indent UCLA Department of Physics and Astronomy \\
\indent Los Angeles, California 90095, USA \\ \\
\noindent September 5, 2006 

\end{document}